\documentstyle[buckow]{article}

\def\beq{\begin{equation}}

\def\eeq{\end{equation}}

\newtheorem{theorem}{Theorem}

\font\elevenmsb=msbm10 scaled 1100

\begin {document}
\begin{raggedleft}
ULB--TH--01/33\\
KUL-TF-2001/23\\ [.5cm]
\end{raggedleft}

\def\titleline{
Comments on duality-symmetric theories\footnote{Talk given by X.B.
at the RTN workshop ``The Quantum Structure of Spacetime and the
Geometric Nature of Fundamental Interactions'', Corfu, 13-20
Sept 2001.}}

\def\authors{

Xavier Bekaert\1ad and Sorin Cucu\2ad }

\def\addresses{

\1ad{\em Physique Th\'eorique et Math\'ematique, Universit\'e
Libre de Bruxelles,}\\ {\em Campus Plaine C.P. 231, B-1050
Bruxelles, Belgium} \\ {\em Fax : +32-2-650.58.24}\\

{\tt xbekaert@ulb.ac.be} \\

\2ad{\em Instituut voor Theoretische Fysica, Katholieke
Universiteit Leuven,}\\ {\em Celestijnenlaan 200D, B-3001 Leuven,
Belgium}\\ {\em Fax. +32-16-32.79.86}\\ {\tt
sorin.cucu@fys.kuleuven.ac.be}

}

\def\abstracttext{We briefly review the basics of electric-magnetic duality
symmetry and their geometric interpretation in the M-theory
context. Then we recall some no-go theorems that prevent a simple
extension of the duality symmetry to non-Abelian gauge theories.}

\large

\makefront

\section{Introduction}

Duality symmetry has a rather old history, going back to the birth
of Maxwell equations. It seems likely that the symmetric role
played by the electric and the magnetic fields in the laws of
electromagnetism, has been one of Maxwell's motivations to modify
Amp\`ere's law by adding the ``displacement current" term
$\frac{\partial {\bf E}}{\partial t}$, which was not sanctioned by
experiment at that time. Maxwell was known to have an aesthetic
appreciation for mathematical structures. In this light, it is
reasonable to endorse the conclusion of Roger Penrose: {\it``It
would seem that the symmetry of these equations and the aesthetic
appeal that this symmetry generated must have played an important
role for Maxwell in his completion of these equations"}
\cite{Penrose}.

As a whole, sourceless Maxwell equations are invariant under the
discrete transformation
\begin{equation}
{\bf E}\rightarrow {\bf B},\quad {\bf B}\rightarrow -{\bf
E}.\label{Z2}
\end{equation}
This is called the electric-magnetic (EM) duality symmetry. The
energy-momentum tensor is also left invariant under this
transformation. This $\mbox{\elevenmsb Z}_2$-duality symmetry
(\ref{Z2}) of sourceless Maxwell theory has been generalized in
various ways. First of all, the discrete duality symmetry extends
to a continuous $SO(2)$ symmetry. Secondly, if we introduce a
$\theta$-term, the EM-duality symmetry group is enhanced from
$SO(2)$ to $SL(2,\mbox{\elevenmsb R})$ group. Thirdly, we can
promote it from an invariance of linear electrodynamics (i.e.
Maxwell theory) to non-linear electrodynamics (e.g. Born-Infeld
theory) \cite{Schr,Zumino}. An other possibility is to introduce
both electric and magnetic sources, in which case the EM-duality
symmetry is preserved if we also ``rotate" the sources in an
appropriate way. Let us mention the important fact that, at the
quantum level, electric and magnetic charges have to satisfy
quantization conditions \cite{Dirac,team}. In the quantum theory,
it follows that the $SL(2,\mbox{\elevenmsb R})$ duality symmetry
group is broken to $SL(2,\mbox{\elevenmsb Z})$ (if $\theta$ is
constrained to vanish, the $SO(2)$ symmetry is reduced to the
previous discrete $\mbox{\elevenmsb Z}_2$ transformation).
Four-dimensional Maxwell theory is an Abelian gauge field theory
with one vector field. The EM-duality symmetry holds in general
for Abelian $p$-form theories in dimension $2(p+1)$ \cite{team}.
In addition, people have considered the case of several Abelian
gauge fields, in which case the duality symmetry is further
enlarged \cite{Zumino}. Recently, the Hodge duality symmetry has
been applied to linearized gravity and higher spin gauge field
theories \cite{Hull}. To conclude this (non-exhaustive) list, let
us add that type IIB superstring theory is expected to have an
$SL(2,\mbox{\elevenmsb Z})$ symmetry, the so-called $S$-duality
\cite{Schwarz}.

The next sections will briefly review some basics of EM-duality
symmetry. Making manifest any symmetry of a theory is always
highly valuable for technical and conceptual issues. In fact, it
is possible to raise EM-duality symmetry to a manifest symmetry of
the action. In section \ref{double} we deal with this question.
Glancing at the huge list of possible extensions of EM-duality
symmetry, it is tempting try to generalize EM-duality to
non-Abelian gauge theories. M-theory even brings some argument to
believe that such a generalization should exist in a way or
another (section \ref{M-}). Unfortunately, some no-go theorems
prevent such an extension if some (not so restrictive) hypotheses
are satisfied. We recall them in the \ref{nogo}th section together
with their hypotheses.

\section{Some basics of EM-duality symmetry}

In order to make higher dimensional generalizations more
transparent, we will reformulate Maxwell equations in terms of
differential forms. To proceed, we define the electric and
magnetic fields as components\footnote{The electric and magnetic
field are the time component of, respectively, the field strength
and its Hodge dual: $E^i=F^{0i}$, $B^i=\frac12
\epsilon^{ijk}F_{jk}$.} of the field strength two-form $F$. The
sourceless Maxwell equations can be rewritten as
\begin{equation}
d \left( \begin{array}{c} F\\ $*$F
\end{array}
\right) = 0\,,\label{Max}
\end{equation}
where $*$ is the Hodge dual. In other words, the field strength
$F$ is a harmonic two-form.

Usually, the Poincar\'e lemma is used to derive from the Bianchi
identity the existence of a potential vector $A$ such that $F=dA$.
Then it is possible to deduce the field equation $d*F=0$ from an
action principle (in all what follows we will omit sources).
Alternatively, the Poincar\'e lemma can be applied to the field
equation to get $*F=d\tilde{A}$. Then the Bianchi identity will be
obtained as the equation of motion of an action depending on
$\tilde{A}$, dual to the Maxwell action.

The sourceless Maxwell equations are manifestly invariant under a
duality rotation:
\begin{equation}
\left( \begin{array}{c} F\\ $*$F
\end{array}
\right) \rightarrow \left( \begin{array}{ll} \cos\alpha &
\sin\alpha\\ -\sin\alpha & \cos\alpha
\end{array}
\right)\left( \begin{array}{c} F\\ $*$F
\end{array}
\right)
 \,.\label{transfo}
\end{equation}
Let us make here the important remark that this duality
transformation is a non-local map in terms of the gauge field $A$.
For an infinitesimal transformation, we have $\delta
A=d^{-1}(*F)\,\delta\alpha$, where $d^{-1}$ stands for the
non-local operator which is the inverse of the differential $d$.

The first obvious generalization is to consider the field strength
$F$ to be a $(p+1)$-form in $D$ spacetime dimensions. Then if we
want $F$ and its dual $*F$ to have the same rank, the dimension is
restricted to $D=2(p+1)$. Naively, the sourceless Maxwell
equations (\ref{Max}) are left invariant by any transformation
(\ref{transfo}) where $R\in GL(2,\mbox{\elevenmsb R})$. But we
have to take into account the relation $*^2F=(-)^p F$ (Minkowskian
spacetime signature). Furthermore, if we ask for the invariance of
the energy-momentum tensor, we get that the global symmetry group
is $SO(2)$ for $D=0\,mod\, 4$, and $\mbox{\elevenmsb Z}_2$ for
$D=2\,mod 4$ \cite{team}. In the following, we will label as
duality-symmetric theories only the first case.

\section{Duality-symmetric theories}\label{double}

Physicists aimed to make EM-duality symmetry manifest in the
action itself. Efforts in this direction have been undertaken also
before (see for instance \cite{z,dt}), but more substantial
results have been achieved during the last decade after the
connections with supergravity and string theory have been pointed
out. As a result one counts different formulations from quadratic
but non-covariant versions \cite{dt,ss}, to quadratic and
covariant actions but with an infinite number of auxiliary fields
\cite{McClain:1990}, or to non-polynomial Lagrangians with
manifest space-time symmetry \cite{pst}, the so-called PST
model\footnote{Using a ``formal" path integral quantization
("formal" in the sense that the possible UV divergences due to the
non-Gaussian character of the integral were not considered) it has
been proved in \cite{Sorin} that the partition function of the PST
and the Maxwell theories are equal. The absence of anomalies and
non-trivial counterterms has been shown recently in \cite{Piguet}.
On this basis, it seems that the PST model can be trusted also at
the quantum level.}. Obviously, it appears that (in any of the
formulations) a high price has to be paid in order to implement
the duality symmetry at the level of the action.

One of the basic ingredients of the duality-symmetric formulation
resides in increasing the number of gauge fields (doubling in that
case) to make global symmetries manifest. At the same time, the
number of gauges symmetries increases in such a way that the
theory possesses the same number of physical degrees of freedom.

Formally, we can solve sourceless Maxwell equations using naively
the Poincar\'e lemma and obtain
\begin{equation}
\left( \begin{array}{c} F\\ $*$F
\end{array}
\right) = d\left( \begin{array}{c} A^1\\ A^2
\end{array}
\right) \equiv \left( \begin{array}{c} F^1\\ F^2
\end{array}
\right)\,.
\end{equation}
The electric and magnetic variables are now on the same footing.
In dimensions $D=0\,mod\,4$, the Hodge square relation imposes for
consistency
\begin{equation}
\epsilon^{ab}F^b=*F^a,\quad(a=1,2).\label{self}
\end{equation} As we can see, the field equation now follow from
the Bianchi identity. Therefore, the idea is simply to obtain this
self-duality equation as e.o.m. derived from an action principle
(this gives some hint that a relation should exist between
EM-duality symmetry and chiral forms, as will be explained in the
next section). This has been achieved in \cite{dt}-\cite{pst}. The
EM-duality rotation (\ref{transfo}) written in terms of the new
variables is
\begin{equation}
\left( \begin{array}{c} A^1\\ A^2
\end{array}
\right) \rightarrow R \left( \begin{array}{c} A^1\\ A^2
\end{array}
\right)
 \,,\quad R\in SO(2).
\end{equation}
The e.o.m. (\ref{self}) is manifestly invariant under this
transformation. Furthermore, a nice feature of duality-symmetric
formulation is that the duality rotation is a local transformation
in terms of the gauge fields $A^a$.

\section{M-theory viewpoint}\label{M-}

New insights on EM-duality symmetry have been provided by
M-theory. Type IIB string theory reduced on a circle is known to
be T-dual to M-theory on a torus. Accordingly, the S-duality
symmetry of the IIB string theory is a consequence of the
invariance of the torus under large diffeomorphisms, the
$SL(2,\mbox{\elevenmsb Z})$ symmetry of the IIB theory being
associated with the modular group of the torus. Thus, in the
M-theory context the S-duality of IIB string theory arises
elegantly from simple geometric arguments. Likewise, the system of
a single M5-brane system provides an appealing geometric
understanding of EM-duality symmetry.

The worldvolume of the M5-brane supports a self-interacting chiral
two-form potential which couples minimally to dyonic strings
located at the intersection of the M5-brane and some M2-branes
ending on it. If the M5-brane is wrapped around the torus, the
T-dual picture in IIB theory is a D3-brane with fundamental
strings ending on it. As a consequence the D3-brane itself is
inert under the modular group $SL(2,\mbox{\elevenmsb Z})$. In
terms of the D3-brane worldvolume theory, this symmetry translates
into the EM-duality symmetry of Abelian Born-Infeld theory
\cite{Berman}. This shed some light on the link between
duality-symmetric theories and chiral forms. Indeed, the e.o.m.
(\ref{self}) in four dimensions finds its origin in the
self-duality equation of the three-form field strength living on
the wrapped M5-brane. A $\mbox{\elevenmsb Z}_2$-duality
transformation then corresponds to the exchange of the two circles
in the compactification from M to IIB theory.

The next step of interest is to consider a system where several,
say $n$, wrapped M5-branes coincide. Unfortunately little is known
about this interacting (2,0) superconformal theory\footnote{Deep
analogies are conjectured to occur for the (4,0) superconformal
theory \cite{Hull2}.}. In the T-dual picture, the M5-branes appear
as a set of coinciding D3-branes. Their dynamics is governed by a
four-dimensional $U(n)$ supersymmetric Dirac-Born-Infeld theory
which, in the weak field limit, is an ordinary $U(n)$ non-Abelian
gauge theory with ${\cal N}=4$ supersymmetry. The determination of
all higher order terms in $\alpha'$ is still an open question but
some progress have been made recently in this direction (see the
talks of M. de Roo, A. Santambrogio and P. Koerber at this
meeting). In any case, from the same arguments as before, the
non-Abelian gauge theory on the coinciding D3-brane worldsheets
should possess an $SL(2,\mbox{\elevenmsb Z})$
symmetry\footnote{This conjectured duality symmetry could be a new
constraint to be imposed in order to derive the full non-Abelian
Born-Infeld action.}.

Turning back to the eleven dimensional picture, the coinciding
D3-branes dynamics suggests that a non-Abelian extension of the
chiral two-form should exist. But the analysis of \cite{X} shows
that the standard Noether procedure will not provide such a theory
as a local deformation of the free one. The interacting theory
could fall outside the scope of perturbative covariant local field
theory \cite{Hull2}. The same could be expected to occur for
$SL(2,\mbox{\elevenmsb Z})$ duality symmetry of four dimensional
non-Abelian vector theory, as we will explain in the following
section.

\section{No-go theorems}\label{nogo}

To conclude, let us consider the sourceless Yang-Mills equations
and look after any duality symmetry property. They read
\begin{equation}
D_A \left( \begin{array}{c} F\\ $*$F
\end{array}
\right) = 0,\label{YM}
\end{equation}
where $A$ is Lie-agebra valued one-form, its curvature is $F= D_A
A=dA+A^2$ and the covariant derivative acts as $D_A =d+[A,\,]$.
Naively, one could think that the Yang-Mills equations (\ref{YM})
are left invariant by the duality rotation (\ref{transfo}).
However, that is not the case since the covariant derivative $D_A$
depends on $A$, which is not inert under the duality rotation.

In the seventies, two no-go theorems \cite{dt,Gu} were found to
prevent such a trivial generalization of EM-duality for Abelian
gauge fields
\begin{theorem}Generically, there is no infinitesimal
transformation $\delta A$ which is able to implement the
infinitesimal duality rotation
\begin{eqnarray}
\delta F=*F\,\delta\alpha\,,\quad\delta
*F=-F\,\delta\alpha.\nonumber
\end{eqnarray}
\end{theorem}
\begin{theorem}There exists one gauge field $A$ solution of $D_A*F=0$
such that there is no gauge field $\tilde A$ which is the ``dual"
of $A$, in the sense that $*F=D_{\tilde{A}}\tilde{A}$.
\end{theorem}
The second no-go theorem teaches us that the Poincar\'e lemma does
not generalize straightforwardly to covariant derivative $D_A$
(Anyway, $D_A$ is not nilpotent since $D^2_A=[F,\,]$). This
prevents a direct application of the scheme of the section
\ref{double} to obtain a duality-symmetric formulation of
Yang-Mills theory.

Of course, no-go theorems have the weakness of their hypotheses.
In consequence, a priori nothing prevents less trivial
generalizations of EM-duality symmetry for non-Abelian gauge
theories. The following no-go theorem\footnote{The proof of the
third theorem is essentially algebraic and is based on a
cohomological reformulation of the Noether procedure for
constructing consistent deformations
\cite{Barnich}.}$^{,}$\footnote{We stress that analogous theorems
hold for any duality-symmetric theories in twice even dimensions.}
further restricts the generalization possibilities
\cite{Bekaert2001}
\begin{theorem}\label{Noway} No consistent, local interactions of a set of free
Abelian vector fields can deform the Abelian gauge transformations
if the local deformed action (free action + interaction terms)
continuously reduces to a sum of free, duality-symmetric,
non-covariant actions in the zero limit for the coupling constant.
\end{theorem}
Since the two main assumptions of the theorem are locality and
continuity of the deformations, in order to escape its conclusion
and to correctly describe non-Abelian gauge field duality, one
should perhaps leave the standard formalism of perturbative local
field theory\footnote{A definition of non-Abelian duality has been
proposed in terms of loop space variables, which are intrinsically
non-local. For recent reviews on this proposal, see for instance
\cite{Tsou}.}.

\section{Acknowledgments}

X.B. is grateful to the organizers for this very enjoyable meeting
and he thanks N. Boulanger and G. Barnich for their remarks
concerning the presentation and content of the talk. X.B. wish
also to thank R. Argurio and G. Bonelli for discussions. This work
was supported in part by the ``Actions de Recherche
Concert{\'e}es" of the ``Direction de la Recherche Scientifique -
Communaut{\'e} Fran{\c c}aise de Belgique", by IISN - Belgium
(convention 4.4505.86) and by the European Commission TMR
programme HPRN-CT-2000-00131, in which X.B. is associated to
Leuven.

%%%%%%%%%%%%%%%%%%%%%%

%%%%%%%%%%%%%%%%%%%%%%

\end{document}